# Effect of activation procedure on Sm-Co-Fe-Zr-B compound for low temperature efficient hydrogen storage

S. S. Makridis[1,2,3,*], Ch. N. Christodoulou[2], E. S. Kikkinides[1] and A. K. Stubos[3]

[1] Department of Mechanical Engineering, University of Western Macedonia, Bakola & Sialvera St., GR-50100, Kozani, Greece

[2] Hystore Technologies Ltd, Ergates Industrial Area, 22[A] Stylli Gonia, CY-2362, Nicosia, Cyprus

[3] Institute of Nuclear Technology and Radiation Protection, NCSR "Demokritos", Ag. Paraskevi, Athens, GR-15310, Greece

# ABSTRACT

The present research work is focused on the effect of activation procedure on the hydrogen absorption-desorption properties of new rare earth – transition metal compound based on $Sm(Co_{0.6}Fe_{0.2}Zr_{0.16}B_{0.04})_{7.5}$ composition. Crystal structure and composition is always connected to the maximum capacity of the intermetallic hydrides. For composite materials the thermodynamic properties of hydrogenation - dehydrogenation procedure are mostly explained through microstructure-microchemistry characteristics. Efficient hydrogen storage is direct connected to the desorbed hydrogen amount. The as hydrogenated material $Sm(Co_{0.6}Fe_{0.2}Zr_{0.16}B_{0.04})_{7.5}$ seems to have in the desorption a pressure plateau below the atmospheric pressure at room temperature while the absorbed hydrogen almost remains in the material having capacity of ~0.8 wt. % at 0.1 MPa - 30 $^{o}$C. After the proper activation procedure, the hydrogenated material desorbs very high amount of hydrogen ~1.9 wt. % at 0.1 MPa - 100 $^{o}$C. Subsequently, the treatment of the composite materials before hydrogenation-dehydrogenation procedures could play a crucial role on the efficiency.

# 1. INTRODUCTION

Hydrogen storage alloys of intermetallic hydride compounds are new functional materials for various applications. Hydrides tank on fuel cell cars is realized because of their compactness and safety for hydrogen energy storage. The hydrogen storage capacity of the alloys is very important for a number of applications like hydrogen tanks, heat pumps, hydride air-conditioners and so on. A comprehensive review of reported intermetallic hydrides it can be found in a database which can be easily accessed via internet at the Sandia Hydrogen Information Center World Web site [1]. Therefore, numerous studies have been done in order to improve the overall properties of these alloys [2-45], which resulted in the discovery of a series of hydrogen storage alloys, including the $AB_5$-type rare earth-based alloys [7], the $AB_2$-type Ti- or Zr-based alloys [8], the Mg-based alloys [9] and the V-based solid solution alloys [10].

New rare earth-transition metal composite multiphase compounds with boron have been recently investigated with interesting hydrogen absorption-desorption characteristics. The effect of boron substitution on microstructure is responsible for the efficient hydrogen storage [11].

In this work we focus on the effect of the pre-hydrogenation (so-called activation) procedure on the $Sm(Co_{0.6}Fe_{0.2}Zr_{0.16}B_{0.04})_{7.5}$ compound. The desorbed amount of hydrogen is associated to the degree of the efficiency for all "dynamic" hydrides.

## 2. EXPERIMENTAL PROCEDURE

The $Sm(Co_{0.6}Fe_{0.2}Zr_{0.16}B_{0.04})_{7.5}$ sample has been prepared by melting the pure metals of Sm, Fe, Co, Cu, Zr and $Fe_3B$ under a voltaic arc. Repeating the melting five times, by inversing the side of the sample each time, has been produced homogeneous sample. Annealing has not been used for higher homogenization.

For the hydrogenation of the material, an outgassing has been performed at 100 $^o$C (3 $^o$C/min) under high vacuum ($10^{-5}$ mbar) for 12 hours (this was deemed sufficient since no mass change has been observed). After cooling down to room temperature the sample container was immersed in a PID temperature controlled oil bath preset at 25 $^o$C or ice/water bath for 0 $^o$C or liquid nitrogen bath for -196 $^o$C. For the isotherms in P-C curve, each equilibrium point $H_2$ was admitted automatically and the pressure was kept constant throughout equilibration by means of motor driven, PID controlled admit and exhaust valves.

A home made "Sieverts apparatus" has been used for the measurements of the desorbed hydrogen. Repeated cycles of the hydrogenation-dehydrogenation procedure have been obtained in order to reach higher desorbed amount of hydrogen. Activation and subsequently hydrogenation has been performed through the use of different conditions of quenching temperature (0 $^o$C, -193 $^o$C) and hydrogen pressure (1.1 MPa, 2.5 MPa). The desorption characteristics of the sample have been examined by means of time and temperature. Dehydrogenation has been obtained at 0.1 MPa (atmospheric pressure). The as hydrogenated sample is denoted for the non activated sample.

## 3. RESULTS AND DISCUSSION

In our previous study [11], has been found that the hydrogen absorption and desorption curves of the as powdered $Sm(Co_{0.6}Fe_{0.2}Zr_{0.16}B_{0.04})_{7.5}$ sample have high hysteresis and very small slope, as shown in Fig. 1.

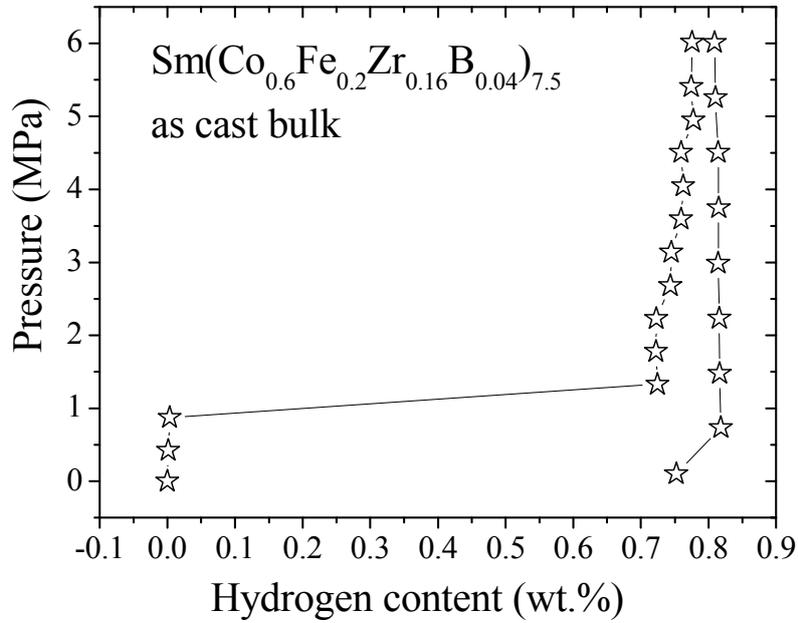

Fig. 1. P - C curve for the $Sm(Co_{0.74}Fe_{0.2}Zr_{0.16}B_{0.04})_{7.5}$ at 26 $^{o}$C [11].

The middle pressure at the pressure plateau in the absorption curve is close to 1.1 MPa (11 bars) while the pressure plateau in the desorption curve is under the atmospheric pressure. Generally, the Sm-Co alloys do not absorb considerable amounts of hydrogen at room temperature [1, 11]. For this as hydrogenated chemical composition, the amount of the absorbed hydrogen is high enough (0.8 wt. %) but unfortunately the desorbed hydrogen at 0.1 MPa (1 bar) is almost zero. Huge changes on the hydrogenation-dehydrogenation characteristics of the compound have been obtained after the proper activation procedure.

The first group of pre-hydrogenation procedures, so-called activation, as shown in Fig. 2, has been used to activate the multiphase compound [11] by charging hydrogen. After putting the sample as a powder in the container, a vacuum has been obtained in the container followed by heating at 100 $^{o}$C (in boiled water bath). At the same temperature a hydrogen pressure of about 1.1 MPa has been applied. Quenching in ice bath has been followed by keeping the pressure constant. The hydrogen charging (hydrogenation) has been realized at the final temperature (30 $^{o}$C) for both charging procedures.

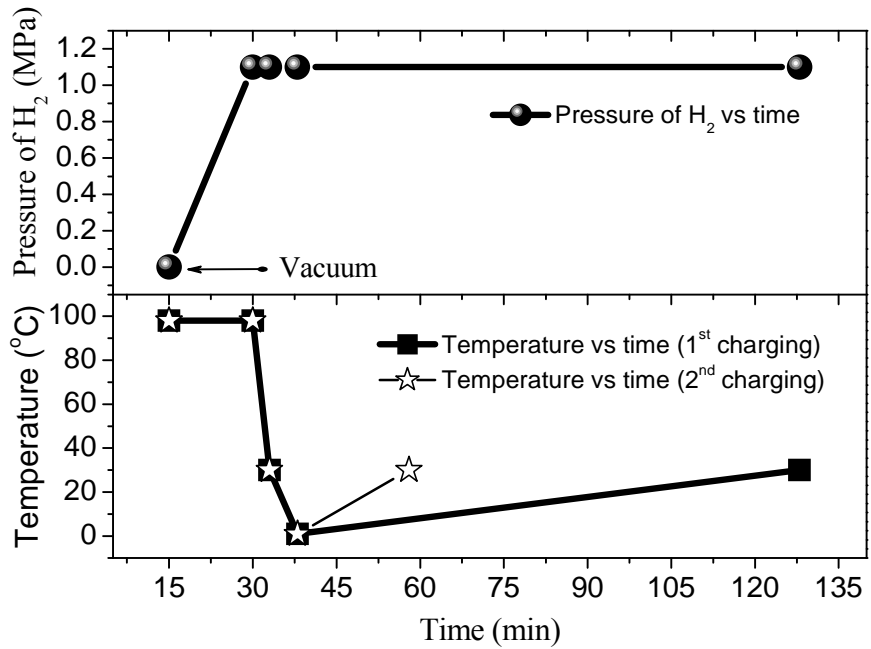

Fig. 2. Activation procedure for charging the as cast $Sm(Co_{0.74}Fe_{0.2}Zr_{0.16}B_{0.04})_{7.5}$ powdered sample at 1.1 MPa-30 °C.

The activated powder, by following the procedure described before, after the first charging desorbs hydrogen amount as high as 70 ml/g (~0.63 wt. % $H_2$) at 0.1 MPa - 60 °C while the desorbed after the second charging has changed to 85 ml/g (~0.76 wt. % $H_2$) at 0.1 MPa - 60 °C, as shown in Fig. 3.

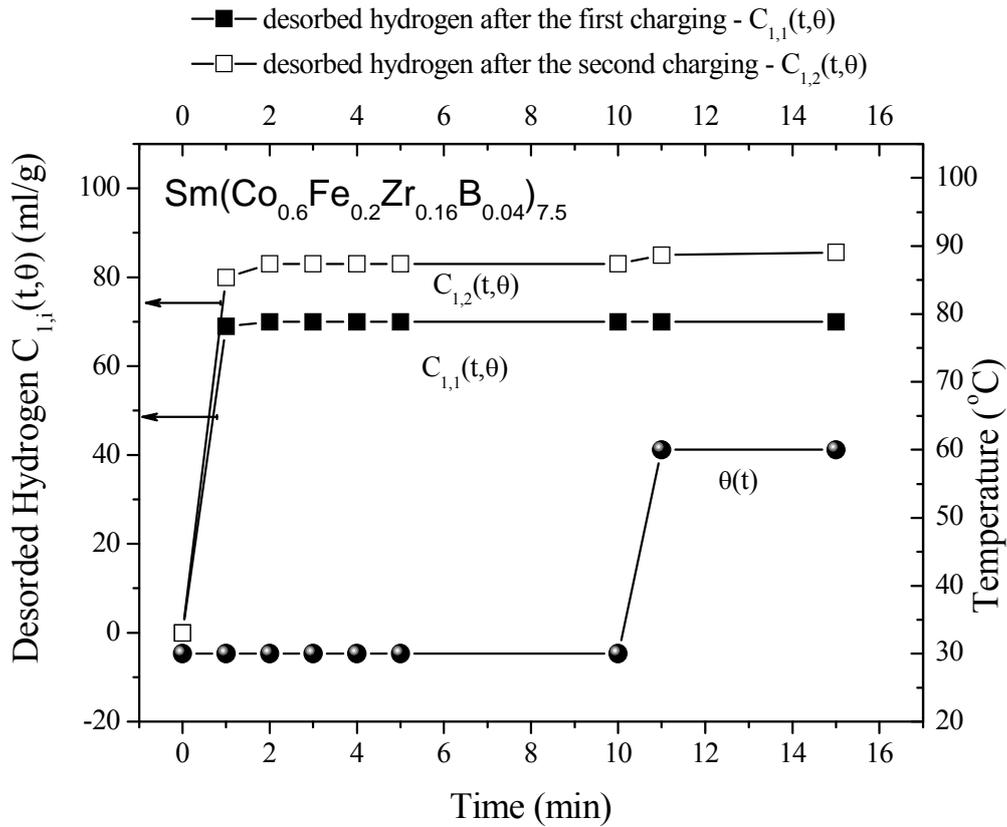

Fig. 3. Desorbed hydrogen $C_{1,i}(t,\theta)$ for the as cast $Sm(Co_{0.74}Fe_{0.2}Zr_{0.16}B_{0.04})_{7.5}$ powdered sample at 30 °C. The subscript, i, represents the activation number.

The second group of activations is shown in Fig. 4. After putting new amount of the sample as a powder in the container, has been performed the same procedure as before but hydrogen pressure of about 2.5 MPa has been applied. Quenching in ice bath has been followed by keeping the pressure constant. The hydrogenation has been realized at the final temperature (30 °C) for all three charging procedures. As was expected, the desorbed amount of hydrogen is much higher the same sample.

After the first hydrogen charging, an almost constant amount of hydrogen has been desorbed from the composite intermetallic compound [11] while temperature did not affect the discharging procedure. By following the second procedure, as shown in Fig. 4, the desorbed hydrogen has been measured to be higher than 60 wt. %. The dehydrogenation properties are shown in Fig. 5.

Activation procedure at 2.5 MPa before hydrogenation at 30 $^{o}$C

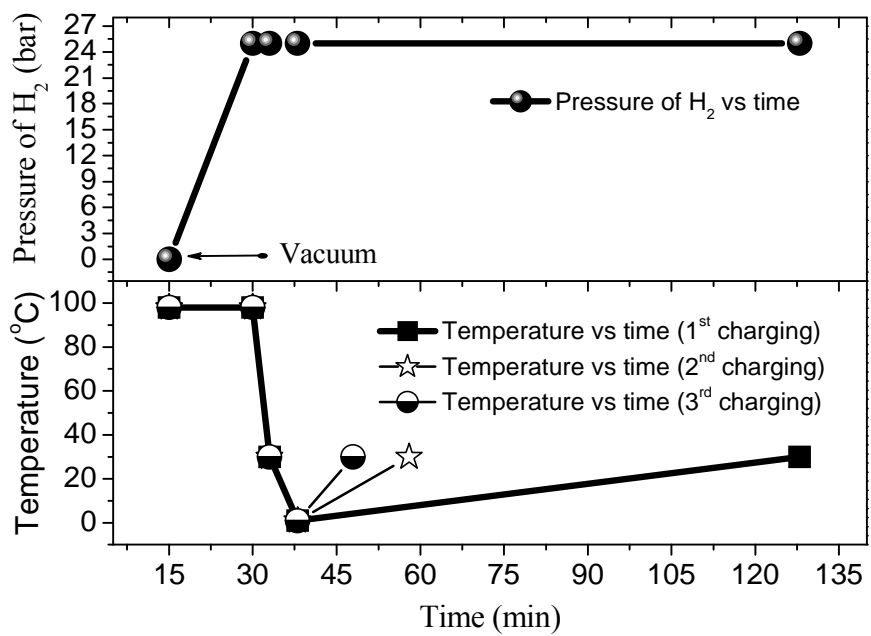

Fig. 4. Activation procedure for charging the as cast Sm(Co$_{0.74}$Fe$_{0.2}$Zr$_{0.16}$B$_{0.04}$)$_{7.5}$ powdered sample at 2.5 MPa-30 $^{o}$C.

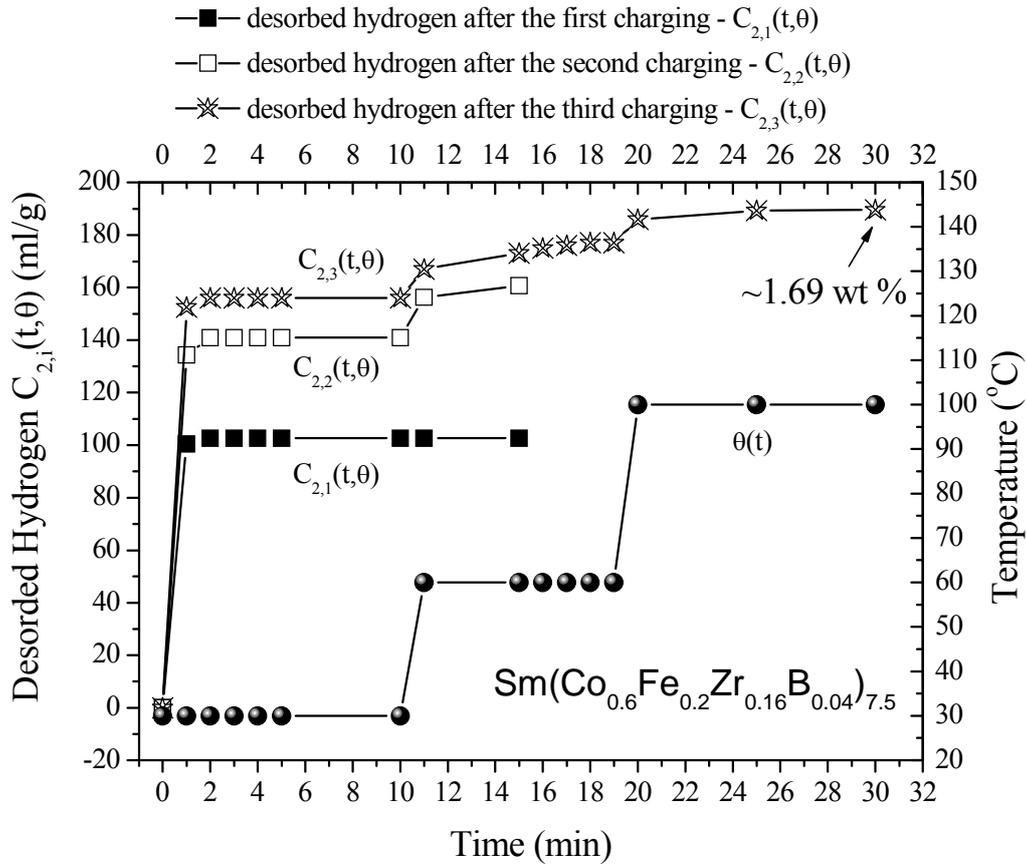

Fig. 5. Desorbed hydrogen $C_{2,i}(t,\theta)$ for the as cast $Sm(Co_{0.74}Fe_{0.2}Zr_{0.16}B_{0.04})_{7.5}$ powdered sample at 30 °C. The subscript, i, represents the activation number.

The non activated powder absorbs almost 0.8 wt % of hydrogen without desorbing it at 26 °C, as shown in Fig. 1. The activated powder, by following the procedure of Fig. 4, desorbs 100 ml/g after the first charging. A high amount of hydrogen has been desorbed after the second charging as high as 141 ml/g (~1.26 wt. % $H_2$) at 30 °C and 161 ml/g (~1.45 wt. % $H_2$) at 60 °C, as shown in Fig. 5. After the third charging, a hydrogen amount of 164 ml/g at 30 °C and 189 ml/g (~1.69 wt. %) has been measured at 100 °C. Therefore, the sample must absorb higher amount of hydrogen than 1.69 wt. %.

Activation procedure is very effective to the degree of the hydrogen-metal reaction. The hydrogenation-dehydrogenation characteristics are modified through changes in microstructure and microchemistry, as always expected after activation procedures. Rare earth - transition metal compounds based on Sm-Co type of alloys have generally very small efficient quantities of hydrogen storage at these conditions of pressure and temperature. While the temperature in the container increases the desorbed hydrogen increases rapidly after the second activation and hydrogen charging. These values that have been measured are completely different on the same sample, compared to that of the Fig. 1. The effect of activation procedure is crucial and needful for maximizing the probability of hydrogen absorption/desorption.

The third group of activation procedures, as shown in Fig. 6, has been performed in order to investigate on the effect of the quenching temperature on the absorbed and subsequently the desorbed amount of the "dynamic" intermetallic hydride $Sm(Co_{0.6}Fe_{0.2}Zr_{0.16}B_{0.04})_{7.5}$. New quantity of powder has been used in the container. After vacuum and heating at 100 °C the sample container, has been applied 2.5 MPa hydrogen pressure. Quenching in liquid nitrogen bath (-196 °C) has been followed by keeping the pressure constant. The hydrogenation has been performed at 30 °C for the four charging procedures.

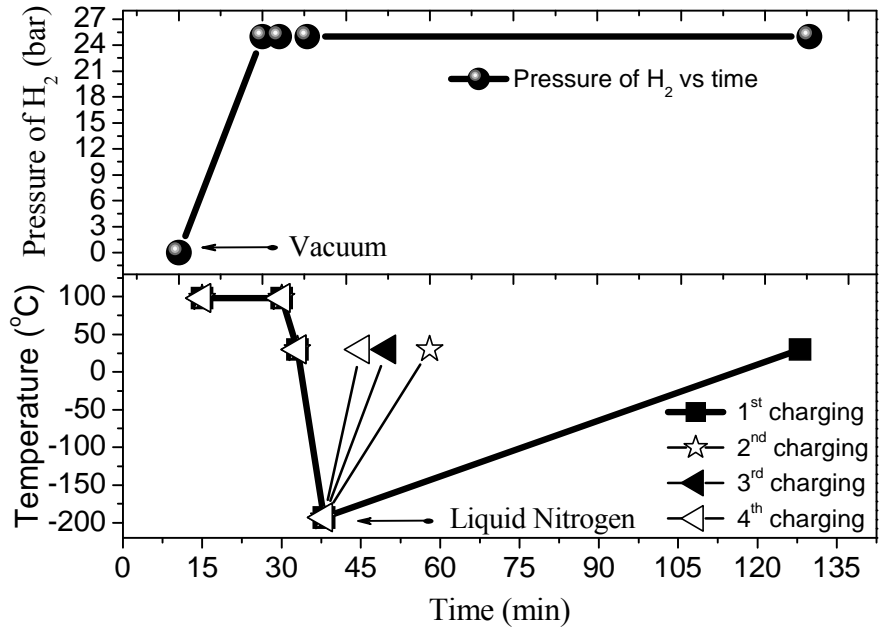

Fig. 6. Activation procedure for charging the as cast $Sm(Co_{0.74}Fe_{0.2}Zr_{0.16}B_{0.04})_{7.5}$ powdered sample at 2.5 MPa-30 °C after quenching in liquid nitrogen bath.

The activated powder, by following the procedure of Fig. 6, desorbs ~120 ml/g after the first hydrogen charging. After the second hydrogen charging, the desorbed amount of hydrogen reaches the ~160 ml/g and ~170 ml/g at 30 °C and 60 °C, respectively, as shown in Fig. 7. After the third charging, a huge amount of efficient hydrogen as high as ~170 ml/g at 30 °C, 190 ml/g at 60 °C, and close to 214 ml/g at 100 °C has been obtained. The last charging has been performed in order to observe any remarkable increase of the desorbed amount of hydrogen.

All thermodynamic characteristics in metal hydrides are dynamic and related to the experimental procedure which is responsible to the hydrogen efficiency.

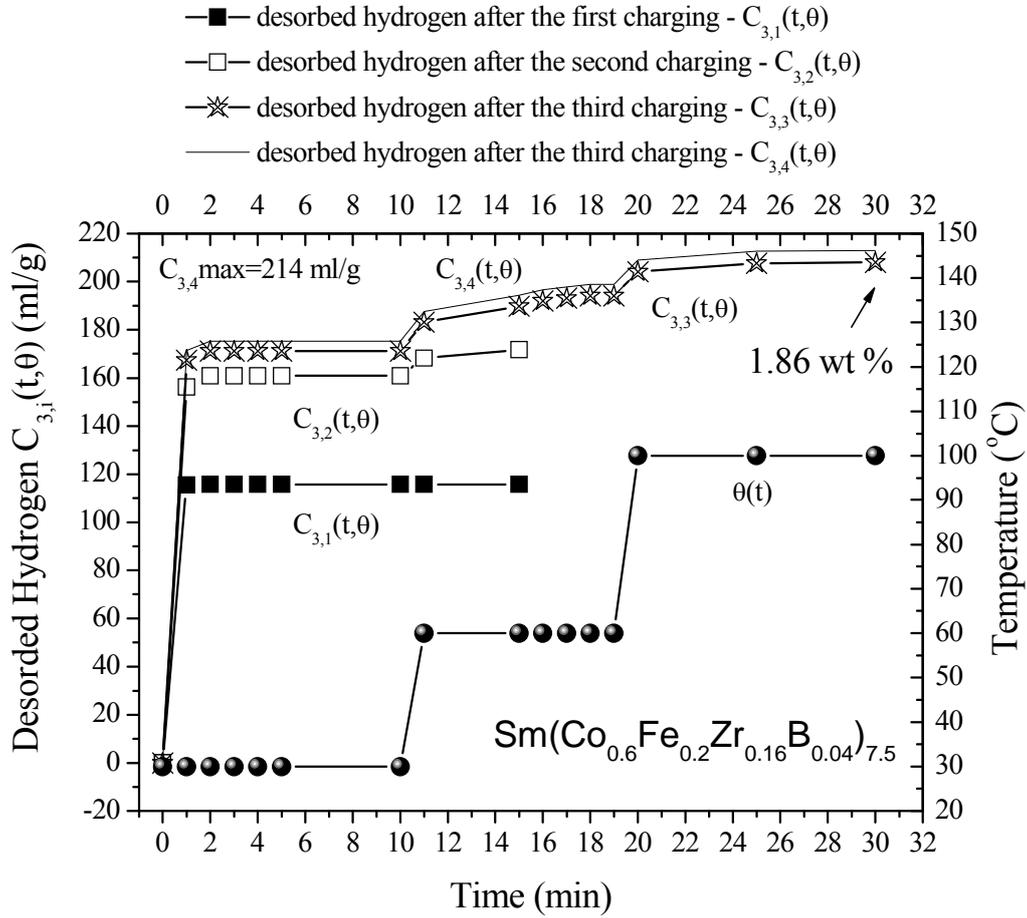

Fig. 7. Desorbed hydrogen $C_{3,i}(t,\theta)$ for the as cast $Sm(Co_{0.74}Fe_{0.2}Zr_{0.16}B_{0.04})_{7.5}$ powdered sample at 30 °C. The subscript, i, represents the activation number.

## 4. CONCLUSIONS

New functional rare earth transition metal intermetallic compound for hydrogen transport has been developed. In this work has been examined the $Sm(Co_{0.6}Fe_{0.2}Zr_{0.16}B_{0.04})_{7.5}$ sample, a new rare earth transition - metal compound, and has been found to absorb and desorb high hydrogen amounts at relatively low temperatures. The as hydrogenated sample keeps almost all amount of hydrogen into the mass. Different groups of activation procedures revealed very different dehydrogenation characteristics. Activation has been found to be very effective to the degree of the hydrogen-metal reaction. Following activation by using quenching in water bath and applying hydrogen pressure of 1.1 MPa at 30 $^o$C has been resulted to desorb hydrogen close to 0.76 wt. % (0.1 MPa - 60 $^o$C). The desorbed amount of hydrogen has been increased to 1.69 wt. % (0.1 MPa - 100 $^o$C) after activation through quenching in water bath and applying hydrogen pressure of 2.5 MPa at 30 $^o$C. The activation procedure with quenching in liquid nitrogen before applying hydrogen pressure of 2.5 MPa at 30 $^o$C, resulted to a huge desorbed amount of hydrogen as high as 1.9 wt. % (0.1 MPa - 100 $^o$C). The kinetics of the hydrogenation and dehydrogenation is fast enough.

The examined efficient hydrogen storage at room or normal temperature and pressure conditions for the $Sm(Co_{0.6}Fe_{0.2}Zr_{0.16}B_{0.04})_{7.5}$ multiphase composite material is highly associated to the activation - hydrogenation procedure. New dynamic hydrogen absorber has been developed for hydrogen energy applications.

.